%% file: ms.tex
\documentclass[camera]{jpaper}
\pdfoutput=1

\usepackage[nocompress]{cite}

\usepackage{soul}

\usepackage{amsmath}

\usepackage[linesnumbered,ruled]{algorithm2e}
\usepackage{graphicx}
\usepackage{algpseudocode}
\usepackage{titlesec}
\algrenewcommand\algorithmiccomment[2][\normalsize]{{#1\hfill\(\triangleright\) #2}}
\usepackage{geometry} \titlespacing*{\section}{0pt}{3pt}{-1pt}
\titlespacing*{\subsection}{0pt}{3pt}{1pt}
\titlespacing*{\subsubsection}{0pt}{1pt}{0pt}

\usepackage{array}
\usepackage{flushend}
\makeatletter
\let\MYcaption\@makecaption
\makeatother

\usepackage[font=footnotesize]{subcaption}

\makeatletter
\let\@makecaption\MYcaption
\makeatother

\usepackage[nolessnomore, italic]{mathastext}
\usepackage[T1]{fontenc}
\usepackage[usenames,dvipsnames,svgnames,table]{xcolor}
\usepackage{multirow}
\usepackage{hhline}
\usepackage[normalem]{ulem}
\usepackage{setspace}
\usepackage{indentfirst}
\usepackage{footmisc}

\usepackage{pifont}

\newif\ifcameraready
\camerareadytrue

\ifcameraready

\else

\fi

\definecolor{amber}{rgb}{1.0, 0.49, 0.0}
\definecolor{darkgreen}{rgb}{0.0, 0.2, 0.13}
\definecolor{darkbyzantium}{rgb}{0.36, 0.22, 0.33}
\definecolor{darkseagreen}{rgb}{0.56, 0.74, 0.56}
\definecolor{darkspringgreen}{rgb}{0.09, 0.45, 0.27}
\definecolor{dollarbill}{rgb}{0.52, 0.73, 0.4}

\usepackage{tikz}
\newcommand*\circled[1]{\tikz[baseline=(char.base)]{
            \node[shape=circle,draw,inner sep=0pt,fill=black, text=white] (char) {#1};}}
\newcommand*\circledWhite[1]{\tikz[baseline=(char.base)]{
            \node[shape=circle,draw,inner sep=0pt,fill=white, text=black] (char) {#1};}}

\include{pythonlisting}

\newcommand{\etal}{\textit{et al.}}

\tikzstyle myBG=[line width=3pt,opacity=1.0]

\definecolor{dred}{rgb}{0.75,0.00, 0.00}
\definecolor{dpink}{rgb}{0.75,0.00, 0.75}
\definecolor{ddpink}{rgb}{1.0, 0.20, 1.0}
\definecolor{dgreen}{rgb}{0.00, 0.75, 0.00}
\definecolor{dblack}{rgb}{0.00, 0.0, 0.00}
\definecolor{dblue}{rgb}{0.00, 0.00, 0.75}
\definecolor{feedb}{rgb}{0.75, 0.00, 0.75}
\colorlet{LightRubineRed}{RubineRed!70!}

\newcommand{\gagan}[1]{{\color{dblack}#1}}
\newcommand{\gagann}[1]{{\color{dblack}#1}}
\newcommand{\gagannn}[1]{{\color{dblack}#1}}

\newcommand{\juan}[1]{{\color{dblack}#1}}
\newcommand{\juang}[1]{{\color{dblack}#1}}
\newcommand{\juangg}[1]{{\color{dblack}#1}}
\newcommand{\juanggg}[1]{{\color{dblack}#1}}

\begin{document}
\bstctlcite{IEEEexample:BSTcontrol} 

 \newcommand{\namePaper}{NERO} 
\title{\namePaper: A Near High-Bandwidth Memory Stencil Accelerator \\for Weather Prediction Modeling \vspace{-0.4cm}}


%


\author{Gagandeep Singh$^{a, b, c}$ \hspace{0.5cm}  Dionysios Diamantopoulos$^c$ \hspace{0.5cm}  Christoph Hagleitner$^c$ \hspace{0.5cm} Juan G{\'o}mez-Luna$^b$ \\  Sander Stuijk$^a$ \hspace{1cm}  Onur Mutlu$^b$ \hspace{1cm}  Henk Corporaal$^a$\\
\vspace{-0.4cm} \normalsize $^a$Eindhoven University of Technology  \hspace{1cm} $^b$ETH Z{\"u}rich   \hspace{1cm}$^c$IBM Research Europe, Zurich\vspace{0.4cm}}



%


\maketitle
\thispagestyle{plain} 
\pagestyle{plain}

\setstretch{1}
\renewcommand{\footnotelayout}{\setstretch{0.9}}







%


\input{abstract}
\input{introduction} 
\input{background}

\input{accelerator.tex}
\input{results}

\input{relatedWork}

\input{conclusion}


\section*{Acknowledgments} 
\label{sec:acknowledgment}
This work was performed in the framework of the Horizon 2020 program for the project ``Near-Memory Computing (NeMeCo)''. It is funded by the European Commission under Marie Sklodowska-Curie Innovative Training Networks European Industrial Doctorate (Project ID: 676240). Special thanks to Florian Auernhammer and Raphael Polig for providing access to IBM systems. We appreciate valuable discussions with Kaan Kara and Ronald Luijten. \gagannn{We would also like to thank Bruno Mesnet and Alexandre Castellane from IBM France for help with the SNAP framework.} This work was partially supported by the H2020 research and innovation programme under grant agreement No 732631, project OPRECOMP.
\juang{We also thank \gagann{Google, Huawei, Intel, Microsoft, SRC, and VMware} for their funding support.}



%


\SetTracking
 [ no ligatures = {f},
 outer kerning = {*,*} ]
 { encoding = * }
 { -40 } 

{

  \let\OLDthebibliography\thebibliography
  \renewcommand\thebibliography[1]{
    \OLDthebibliography{#1}
    \setlength{\parskip}{0pt}
    \setlength{\itemsep}{0pt}
  }
  \bibliographystyle{IEEEtranS}
  \bibliography{ref}
}

\end{document}

%% file: abstract.tex
\begin{abstract}
Ongoing climate change calls for \gagan{fast and accurate} 
weather and climate \gagan{modeling}. However, \gagan{when solving large-scale \gagann{weather prediction}
simulations,} state-of-the-art CPU and GPU implementations suffer from limited performance and high energy consumption. \juan{These implementations are \gagan{dominated by} complex irregular memory access patterns and low arithmetic intensity \gagan{that} pose fundamental challenge\gagannn{s} to acceleration}. To overcome these challenges, we propose \gagan{and evaluate} the use of near-memory acceleration using a reconfigurable fabric with high-bandwidth memory (HBM). \gagan{We focus on} \juan{compound stencils \gagan{that} are fundamental kernels} in weather prediction models. \gagan{By using} high-level synthesis techniques, \gagan{we} develop \juang{\namePaper,} an \juang{FPGA+HBM-based} accelerator connected through IBM CAPI2 (Coherent Accelerator Processor Interface) to an IBM POWER9 host system. \gagan{Our} experimental results show that \juang{\namePaper} outperforms a 16-core POWER9 \gagann{system} by $4.2\times$ and $8.3\times$ when running two \juanggg{different} compound stencil kernels. \juang{\namePaper} \gagann{reduces the energy consumption by} $22\times$ and $29\times$ \juanggg{for the same two kernels \gagannn{over the POWER9 system}} \gagann{with an energy efficiency of 1.5 GFLOPS/Watt and 17.3 GFLOPS/Watt}. \gagan{We conclude that} employing near-memory acceleration solutions for weather prediction \gagan{modeling}~\gagan{is~promising} \juanggg{as a means to achieve \gagannn{both} high performance and \gagannn{high} energy efficiency}.
        
\end{abstract}

%% file: introduction.tex
\section{Introduction} 
{Accurate weather prediction using detailed weather models is essential to \gagan{make} weather-dependent decisions in a timely manner.}
The Consortium for Small-Scale Modeling (COSMO)~\cite{doms1999nonhydrostatic} 
\juang{built} one such weather model 
to meet the high-resolution forecasting requirements of weather services. The COSMO model is a non-hydrostatic atmospheric prediction model currently being used by a dozen nations for meteorological purposes and research applications. 

The central part of the COSMO model (\juan{called \emph{dynamical core} or \emph{dycore}}) solves the Euler equations on a curvilinear grid and applies implicit discretization \gagann{(i.e., parameters are dependent on each other at the same time instance~\cite{bonaventura2000semi})} in the vertical dimension and 
explicit discretization \gagan{(i.e., 
\juanggg{a solution is dependent on the previous system state}~\cite{bonaventura2000semi})} in the horizontal dimension. The use of different discretizations leads to three computational patterns~\cite{cosmo_knl}: \juang{1)} horizontal stencils, \juang{2)} tridiagonal solvers in the vertical dimension, and \juang{3)} point-wise computation. These computational kernels are compound stencil kernels that operate on a three-dimensional grid~\cite{gysi2015modesto}.  
\emph{Vertical advection} (\gagan{\texttt{vadvc}}) and  \emph{horizontal diffusion} (\gagan{\texttt{hdiff}})  are such compound kernels found in the \emph{dycore} of the COSMO \gagannn{weather prediction} model. {These kernels} are representative \gagan{of} the data access {patterns and algorithmic} complexity of the entire COSMO model. \gagan{They} are similar to the kernels used in other weather and climate models~\cite{kehler2016high,neale2010description,doi:10.1175/WAF-D-17-0097.1}.  Their performance is dominated by memory-bound operations with unique irregular memory access patterns 
\juang{and} low arithmetic intensity that often results in $<$10\% sustained floating-point performance on current CPU-based systems~\cite{chris}. 

\begin{figure}[h]
  \centering
  \includegraphics[width=0.5\textwidth,trim={1cm 0.8cm 1.5cm 0.6cm},clip]{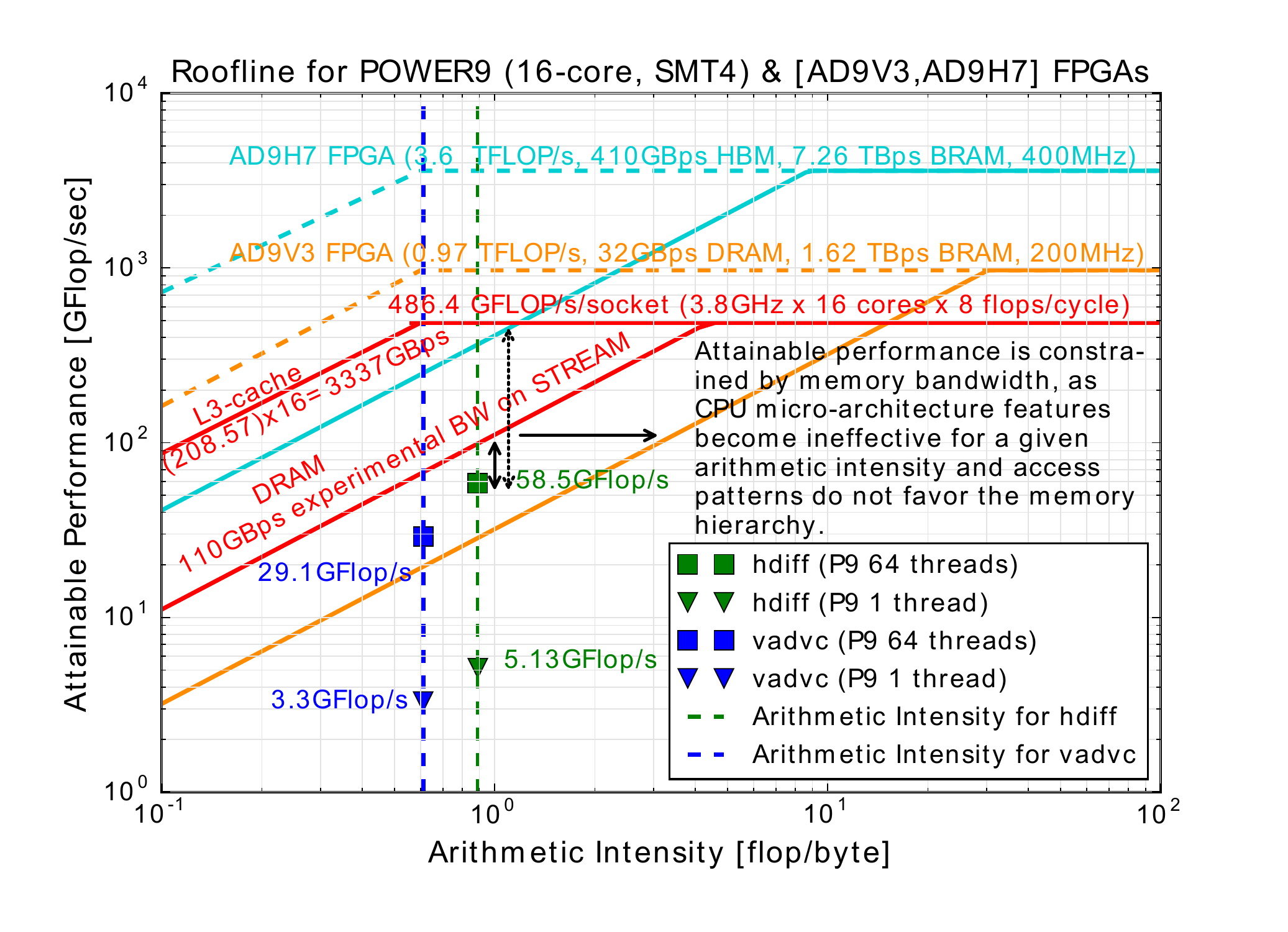}
    \caption{Roofline~\cite{williams2009roofline} for POWER9 (1-socket) showing vertical advection {(\texttt{vadvc})} and horizontal diffusion {(\texttt{hdiff})} kernels for single-thread and 64-thread 
    implementations. 
    \juang{The plot shows also the rooflines} 
    of the FPGAs used in our work.}
 \label{fig:roofline}
\end{figure}

Figure~\ref{fig:roofline} shows the roofline plot\gagan{~\cite{williams2009roofline}} for {an} IBM 16-core POWER9 CPU (IC922).\footnote{IBM and POWER9 are registered trademarks or common law marks of International Business Machines Corp., registered in many jurisdictions worldwide. Other product and service names might be trademarks of IBM or other companies.} After optimizing the \texttt{vadvc} and \texttt{hdiff} kernels for the POWER architecture by following the approach in~\cite{xu2018performance}, 
\juan{they} achieve {29.1}~GFLOP/s and 58.5~GFLOP/s, \juan{respectively}, \juang{for 64 threads}. {Our roofline analysis 
\juang{indicates} that these kernels are constrained by the host DRAM bandwidth.} 
\juan{Their} low arithmetic intensity limits 
\juang{their performance, which is one order of magnitude smaller than the peak performance,} and results in a fundamental memory bottleneck that 
standard CPU-based optimization techniques \juan{cannot overcome}.

\juang{In this work, our goal is to overcome the memory} {bottleneck \juang{of weather \gagan{prediction} kernels} by \gagan{exploiting near-memory \gagann{computation} capability on} FPGA accelerators with high-bandwidth memory (HBM)~\cite{6757501,hbm,lee2016smla} \gagan{that are attached} \juang{to the host CPU}. 
Figure~\ref{fig:roofline} shows the roofline model\gagann{s} of the two FPGA cards \juang{(AD9V3~\gagan{\cite{ad9v3}} and AD9H7~\gagan{\cite{ad9h7}})} used in this work. 
FPGAs \juang{can} \gagan{handle} 
irregular memory access patterns \juang{efficiently} and 
offer significantly 
\juang{higher} memory bandwidth \juang{than the host CPU with} 
their on-chip URAMs \gagan{(UltraRAM)}, BRAMs \gagan{(block RAM)}, and \juang{off-chip} HBM (\gagan{high-bandwidth memory} for the AD9H7 card).} 
However, taking full advantage of 
FPGAs for accelerating a workload is not a trivial task. To compensate \gagan{for} the higher clock frequency of the \gagan{baseline} CPUs, 
\gagan{our} FPGAs must exploit at least one order of magnitude more parallelism in a target workload. 
\juang{This is challenging, as it requires sufficient FPGA programming skills to \gagan{map the workload and} optimize the design for the FPGA microarchitecture.}

\juang{Modern FPGA boards deploy} 
\juan{new} cache-coherent interconnects, such as 
IBM CAPI~\cite{openCAPI}, Cache Coherent Interconnect for Accelerators
(CCIX)~\cite{benton2017ccix}, and Compute Express Link (CXL)~\cite{sharma2019compute}, \juang{which} allow tight integration of FPGAs with CPUs at 
high bidirectional bandwidth \juang{(\gagan{on} the order of tens of GB/s)}. However, memory-bound applications on FPGAs are limited by \juang{the relatively} low DDR4 bandwidth (72 GB/s 
\gagan{for four independent dual-rank DIMM interfaces}\gagann{~\cite{VCU}}). 
\juan{To overcome this limitation,} FPGA vendors have started offering devices \juan{equipped} with HBM~\cite{xilinx_utlra,hbm_specs,intel_altera,lee2016smla} 
\juan{with} a theoretical peak bandwidth of 410~GB/s. 
\juang{HBM-equipped} FPGAs have the potential to \gagannn{reduce} the \gagannn{memory} bandwidth bottleneck, {but} a study of their advantages for 
\juang{real-world memory-bound} applications 
is still~missing.


\juan{We aim} to answer the following \juan{research question}: \textbf{Can FPGA-based \gagan{accelerators} with HBM \gagan{mitigate} the performance bottleneck of \juang{memory-bound} compound weather \gagan{prediction} kernels 
in an energy efficient way?} As an answer to this question\gagann{,} we present \namePaper, a \underline{ne}ar-HBM accelerator for weathe\underline{r} predicti\underline{o}n. 
\juan{We design and implement} \namePaper~
\juan{on an FPGA with HBM to} optimize two 
kernels (vertical advection and horizontal diffusion), 
\juan{which notably} represent 
\juan{the} spectrum of {computational diversity} found in the COSMO \gagann{weather prediction} application. 
\juang{We co-design a hardware-software framework and provide an optimized API to interface efficiently with the rest of the COSMO model, which runs on the CPU}.
Our \gagan{FPGA-based} solution \gagan{for \texttt{hdiff} and \texttt{vadvc}} leads to 
\gagan{performance improvements of $4.2\times$ and $8.3\times$ and}
\juang{energy reductions} of $22\times$ and $29\times$, \juanggg{respectively,} \juang{with respect to optimized CPU implementations~\cite{xu2018performance}}. 

\gagan{T}he major contributions of this paper are \juang{as follows}:
\begin{itemize}
\item \juang{We perform \gagan{a detailed} roofline analysis to show that representative weather \gagan{prediction} kernels are constrained by memory bandwidth on \gagannn{state-of-the-art} CPU systems.}

\item {We propose \namePaper, \gagan{the first} near-HBM FPGA-based accelerator} for representative kernels from a real-world weather prediction application.

\item \juang{We optimize \namePaper~with} a data-centric caching scheme with precision-optimized tiling for a heterogeneous memory hierarchy \gagan{(consisting of URAM, BRAM, and HBM)}.

\item 
\juan{We evaluate the performance and energy consumption of our accelerator and perform a scalability analysis.} 
We show that an FPGA+HBM-based design 
outperforms a complete 16-core POWER9 \gagann{system}~(
\juang{running} 64~threads)~by~$4.2\times$ for 
\juang{the} {vertical advection} \juanggg{(\texttt{vadvc})} and $8.3\times$ for 
\juang{the} {horizontal diffusion} \juanggg{(\texttt{hdiff})} kernels with energy \gagann{reductions of $22\times$ and $29\times$, respectively. Our design provides an energy efficiency of 1.5 GLOPS/Watt and 17.3 GFLOPS/Watt for \texttt{vadvc} and \texttt{hdiff} kernels,} \juanggg{respectively}.

\end{itemize}


%% file: background.tex
\section{Background} 
\label{sec:background}
\juang{In this section, we first provide} 
an overview of the \juang{\texttt{vadvc} and \texttt{hdiff}} {compound stencils, 
\juang{which represent} a large fraction of the overall computational load of the COSMO \gagannn{weather prediction} model}. 
Second, we introduce the CAPI SNAP \gagan{(Storage, Network, and Analytics Programming)} framework\footnote{https://github.com/open-power/snap} that we use to connect our \namePaper~accelerator to an IBM POWER9~system.

\subsection{\juangg{Representative} COSMO Stencils}
A stencil operation updates values in a structured multidimensional grid based on the values of a fixed local neighborhood of grid points. 
Vertical advection (\texttt{vadvc}) and horizontal diffusion (\texttt{hdiff}) from the COSMO model are two such compound \juang{stencil} kernels, which represent the typical code patterns found in the \emph{dycore} of COSMO.  
Algorithm~\ref{algo:hdiffKernel} shows \gagan{the} pseudo-code for \texttt{vadvc} and \texttt{hdiff}  kernels. 
The horizontal diffusion kernel iterates over a 3D grid performing \textit{Laplacian} and \textit{flux} to calculate different grid points. Vertical advection has a higher degree of complexity since it uses the Thomas algorithm~\cite{thomas} to solve a tri-diagonal matrix of the velocity field along the vertical axis. Unlike the \gagan{conventional} stencil kernels, vertical advection has dependencies in the vertical direction, which leads to limited available parallelism.

\input{algorithms/cosmo.tex}

Such compound kernels are dominated by memory-bound operations with complex memory access patterns and low arithmetic intensity. This poses a fundamental challenge to acceleration. 
\juanggg{CPU implementations of these kernels~\cite{xu2018performance} suffer from limited data locality and inefficient memory usage, as our roofline analysis in Figure~\ref{fig:roofline} exposes}.

\subsection{CAPI SNAP \juang{Framework}}
The OpenPOWER Foundation Accelerator Workgroup~\cite{open_power} created the CAPI SNAP framework, 
\juanggg{an open-source environment for FPGA programming productivity}. 
CAPI SNAP \gagan{provides two key benefits}~\cite{wenzel2018getting}: (i) it enables an improved developer productivity for FPGA acceleration and eas\gagan{es the} use of CAPI's cache-coheren\gagan{ce mechanism}, and (ii) it places 
\juanggg{FPGA-}accelerated compute engines, 
\juanggg{also known as} FPGA \textit{actions}, closer to \gagan{relevant} data to achieve better performance. SNAP provides a simple API to invoke an accelerated \textit{action}, and also provides programming methods to \gagan{instantiate} customized accelerated \textit{actions} on the FPGA side. These accelerated \textit{actions} can be specified in C/C++ code that is then compiled to the FPGA target using the Xilinx Vivado High-Level Synthesis (HLS) tool~\cite{hls}.

%% file: algorithms/cosmo.tex
\begin{algorithm}[h]
\scriptsize
\SetAlgoLined
\DontPrintSemicolon
\SetNoFillComment
\caption{Pseudo-code for vertical advection and horizontal diffusion \gagan{kernels}
used by the COSMO~\cite{doms1999nonhydrostatic} \gagannn{weather prediction} model.}
\label{algo:hdiffKernel}
\SetKwFunction{FMain}{verticalAdvection}
   \SetKwFunction{FTest}{forwardSweep}
    \SetKwFunction{FBack}{backwardSweep}
\SetKwProg{Fn}{Function}{}{end}
  
  \Fn{\FMain{float* ccol,
                float* dcol,
                float* wcon,
                float* ustage,
                float* upos,
                float* utens,
                float* utensstage
           }}{
       
             \For{$c\gets2$ \KwTo $column-2$}{
                    \For{$r\gets2$ \KwTo row-2}{
                    
               \Fn{\FTest{float* ccol,
                float* dcol,
                float* wcon,
                float* ustage,
                float* upos,
                float* utens,
                float* utensstage}} { 
                    \For{$d\gets1$ \KwTo $depth$}{
                       \tcc*[l]{forward sweep calculation}
                       }
                     }   
                       
                \Fn{\FBack{float* ccol,
                float* dcol,
                float* wcon,
                float* ustage,
                float* upos,
                float* utens,
                float* utensstage}} { 
                    \For{$d\gets depth-1$ \KwTo$ 1$}{
                       \tcc*[l]{backward sweep calculation}}}
                    }
                  }

}


\hrule
\SetKwFunction{FMain}{horizontalDiffusion}
\SetKwProg{Fn}{Function}{}{end}
  
  \Fn{\FMain{float* src, float* dst}}{
       \For{$d\gets1$ \KwTo $depth$}{
             \For{$c\gets2$ \KwTo $column-2$}{
                    \For{$r\gets2$ \KwTo row-2}{
                        \tcc*[l]{\gagan{L}aplacian calculat\gagan{ion}}
                        $lap_{CR}=laplaceCalculate(c,r)$
                        \tcc{row-laplacian}
                        $lap_{CRm}=laplaceCalculate(c,r-1)$\;      
                        $lap_{CRp}=laplaceCalculate(c,r+1)$
                        \tcc{column-laplacian}
                        $lap_{CmR}=laplaceCalculate(c-1,r)$\;
                        $lap_{CpR}=laplaceCalculate(c+1,r)$
                        \tcc{column-flux calculat\gagan{ion}}\
                        $flux_{C} = lap_{CpR} - lap_{CR}$\;
                        $flux_{Cm} = lap_{CR} - lap_{CmR}$\;
                        \tcc{row-flux calculat\gagan{ion}}
                        $flux_{R} = lap_{CRp} - lap_{CR}$\;
                        $flux_{Rm} = lap_{CR} - lap_{CmR}$\;
                        \tcc{output calculat\gagan{ion}}
                        $dest[d][c][r] = src[d][c][r] -\linebreak
                            c1 * (flux_{CR}- flux_{CmR}) + (flux_{CR}- flux_{CRm})$
                        }
            
                }
   
         }
}
\end{algorithm}

%% file: accelerator.tex
\section{Design Methodology}
\label{sec:design}
\subsection{NERO, A Near HBM Weather Prediction Accelerator}
The low arithmetic intensity of real-world weather \gagan{prediction} kernels limits the attainable performance on current multi-core systems. This sub-optimal performance is due to \gagan{the kernels'} complex memory \juanggg{access} patterns and their inefficiency in exploiting a rigid cache hierarchy, \gagan{as quantified in \juangg{the} roofline plot in} Figure~\ref{fig:roofline}. 
These kernels \gagannn{cannot} fully utilize the available memory bandwidth, which leads to 
high 
\juanggg{data movement overheads} in terms of latency and energy consumption. We address 
\juanggg{these inefficiencies} by developing an architecture \gagan{that} combines fewer off-chip data \gagan{accesses} with higher throughput for the loaded data. \gagan{To this end}, our accelerator design \gagan{takes} a data-centric approach~\cite{mutlu2019,ghose2019processing,teserract,singh2019near,NAPEL,hsieh2016accelerating,7551394,ahn2015pim,googleWorkloads,kim2018grim} \gagan{that exploits} near high-bandwidth memory acceleration.

\begin{figure}[t]
  \centering
  \includegraphics[width=0.8\linewidth,trim={0.3cm 0.2cm 0cm 0.2cm},clip]{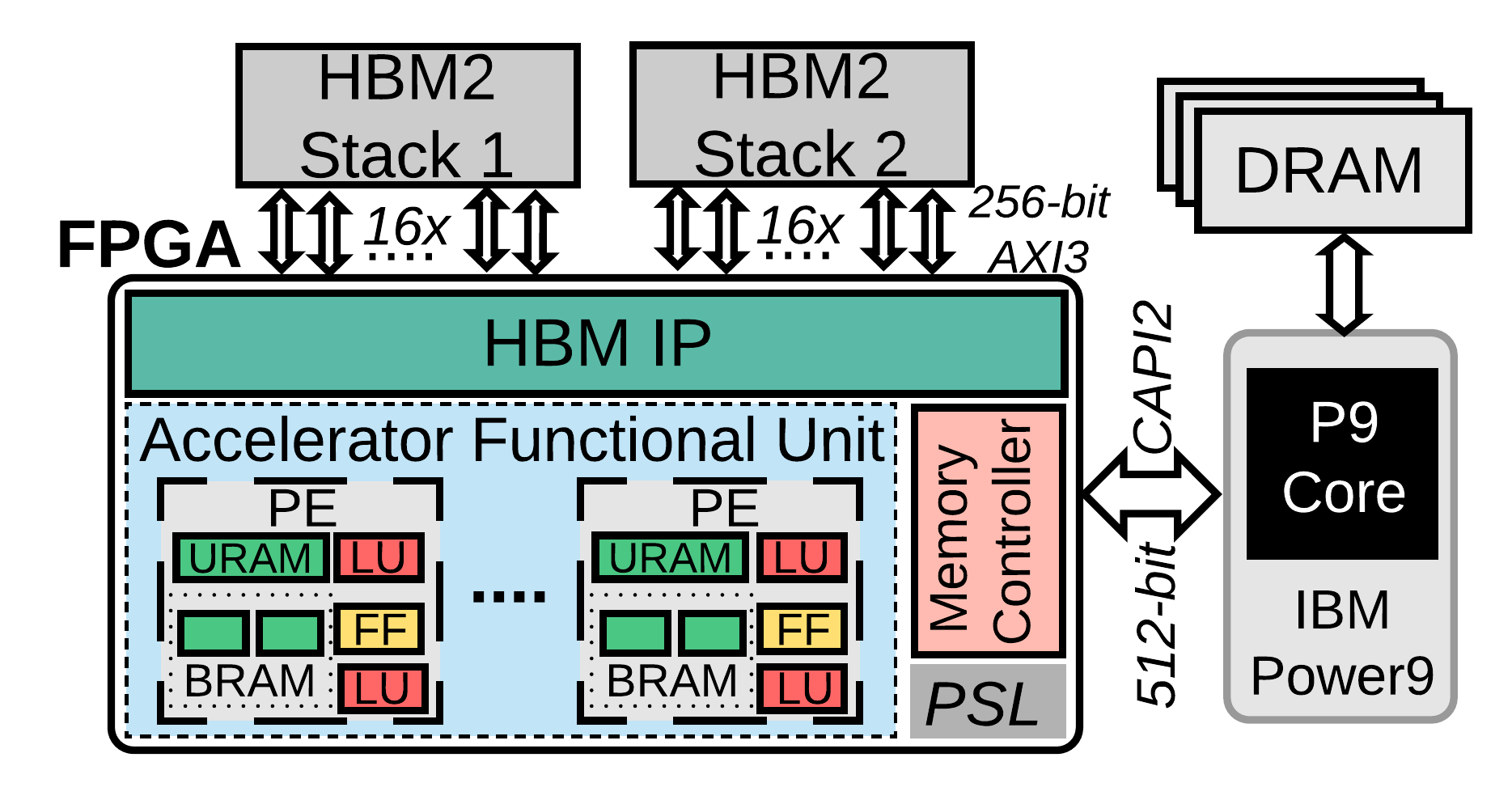} 
  \caption{Heterogeneous platform with an IBM POWER9 system connected to an HBM-based FPGA board via~CAPI2.
 \label{fig:system}}
 \end{figure}

\begin{figure*}[t]
\begin{subfigure}[t]{0.7\textwidth}
  \includegraphics[width=0.95\linewidth,trim={0.5cm 0.5cm 0cm 0cm},clip]{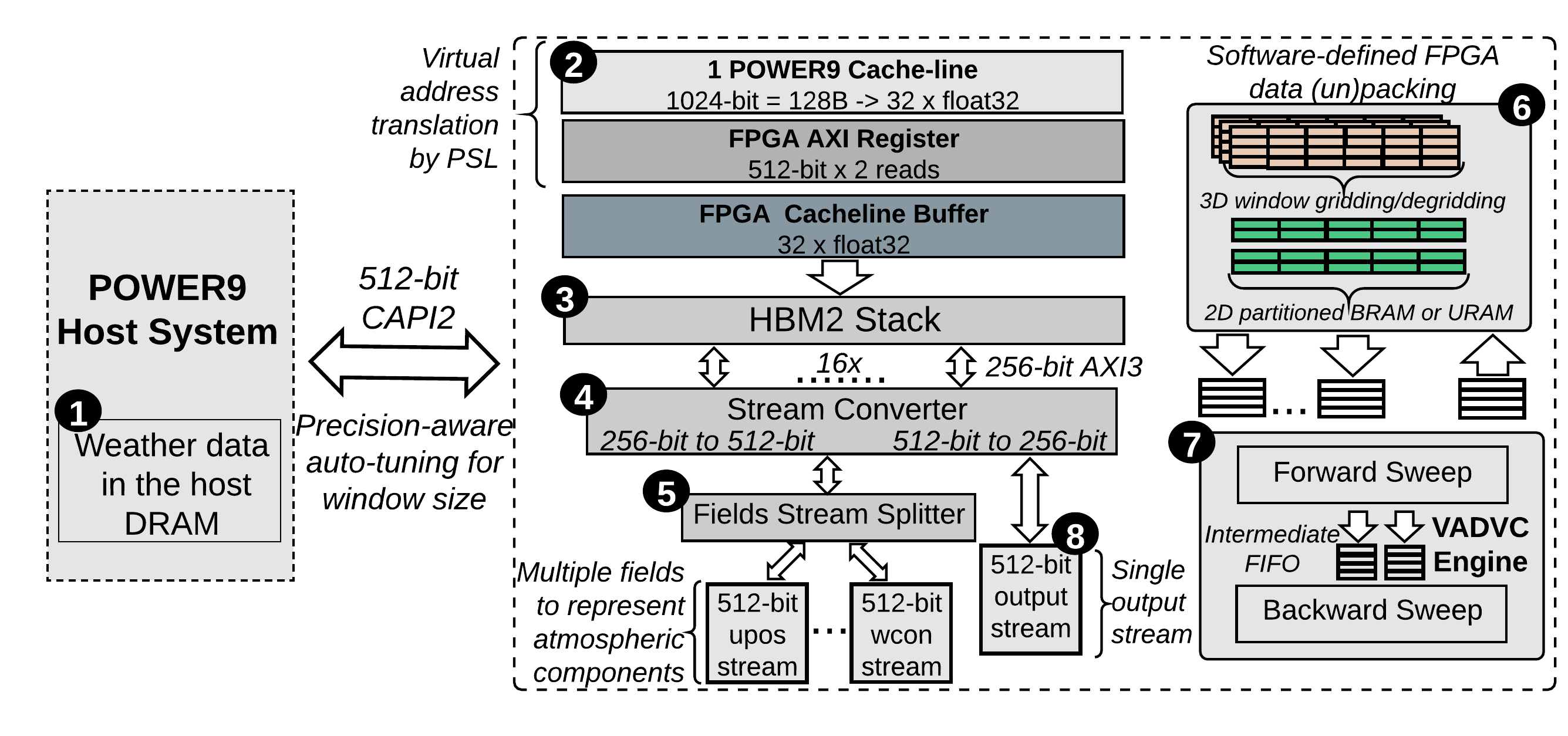} 
   \caption{
  \label{fig:complete_flow}}
\end{subfigure}%
\begin{subfigure}[t]{0.3\textwidth}
  \includegraphics[width=0.95\linewidth,trim={0cm 0cm 0cm 0cm},clip]{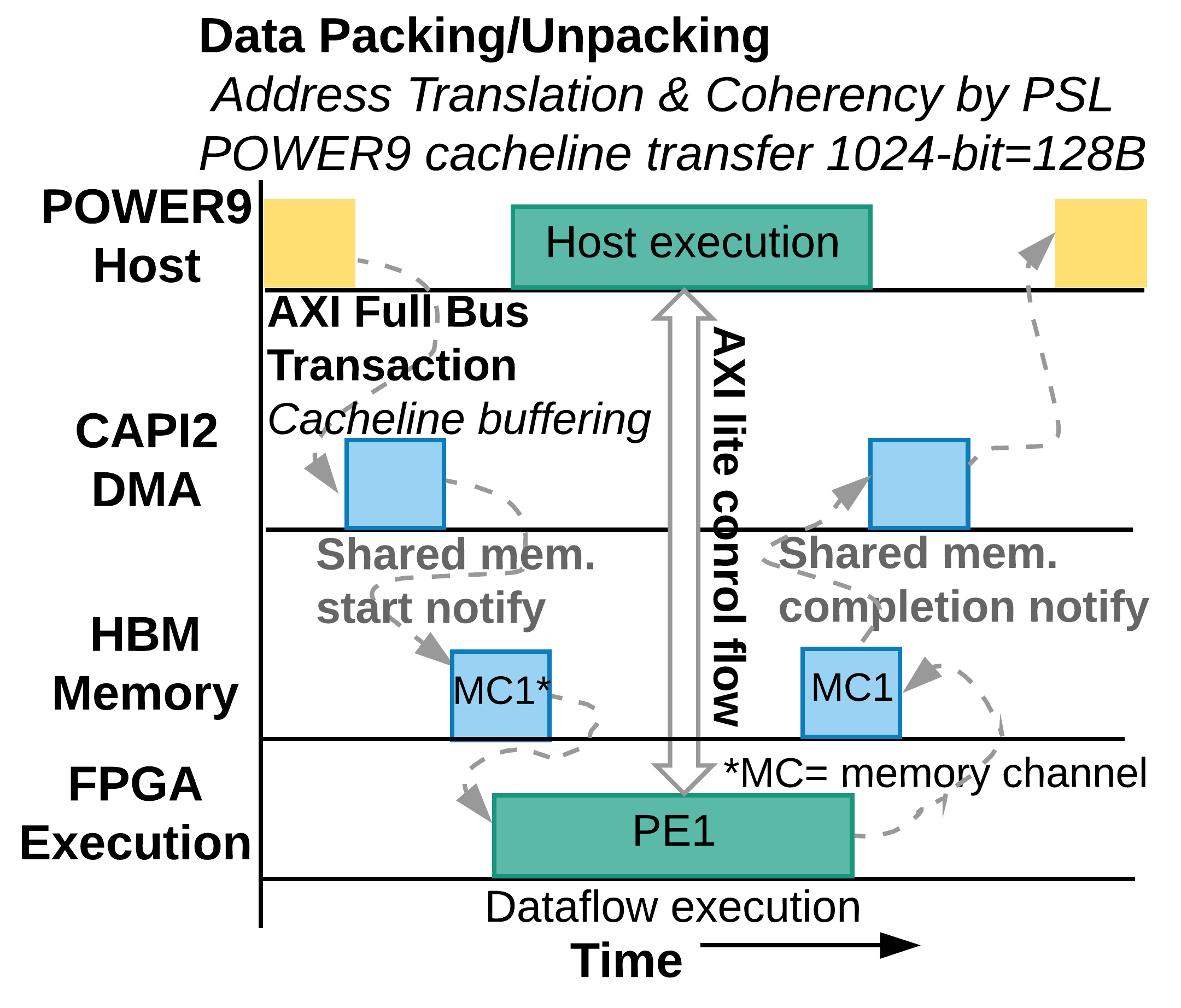} 
   \caption{
  \label{fig:execution}}
\end{subfigure}
\caption[Two numerical solutions]{(a) Architecture overview of~\namePaper~with data flow sequence from the host DRAM to the on-board FPGA memory via POWER9 cachelines. We {depict a} single {processing element (PE)} fetching data from a dedicated HBM port. The number of HBM ports scales linearly with the number of {PEs}. Heterogeneous partitioning of on-chip memory blocks \juangg{reduces} read \juangg{and} write latencies across the FPGA memory hierarchy. (b)  Execution timeline with data flow sequence from the  host  DRAM  to the onboard  FPGA  memory.\label{fig:complete_hbm_flow}}
\end{figure*}

Figure~\ref{fig:system} shows a high-level overview of our integrated system. \gagan{An HBM-based} FPGA is connected to a server system based on an IBM POWER9 processor using the \juangg{Coherent Accelerator Processor Interface version} 2 (CAPI2). 
The FPGA consists of two HBM stacks\footnote{In this work, we enable only a single stack based on our resource and power consumption analysis \gagan{for \gagannn{the} \texttt{vadvc} kernel.}}, each with 16 \emph{pseudo-memory channels}~\cite{axi_hbm}. \gagan{A} channel is exposed to the FPGA as a 256-bit wide port, and in total, 
\juanggg{the FPGA} has 32 such ports. The HBM IP provides 8 memory controllers \juanggg{(per stack)} to handle the data transfer to and from the HBM memory ports. Our design consists of an \emph{accelerator functional unit} (AFU) that interacts with the host system through the power service layer (PSL), which is the CAPI endpoint on the FPGA. An AFU comprises of multiple \emph{processing elements} (PEs) that perform compound stencil computation. Figure~\ref{fig:complete_flow} shows the architecture overview of \namePaper. As vertical advection is the most complex kernel, we depict \gagan{our architecture design flow for} vertical advection. We use a similar design for the \gagan{horizontal diffusion} kernel.

The weather data, based on the atmospheric model resolution grid, is stored in the DRAM of the host system \gagan{(\circled{1} in Figure~\ref{fig:complete_flow})}.  We employ \juangg{the} double buffering technique  between the CPU and the FPGA to hide the PCIe \gagan{(Peripheral Component Interconnect Express~\cite{pcie})} \gagannn{transfer} {latency}.  {By configuring a buffer of 64 cache lines, between the AXI4 interface of CAPI2/PSL and the AFU, we can reach the theoretical peak bandwidth of CAPI2/PCIe (\juanggg{i.e.,} 16 GB/s).} We create a specialized memory hierarchy from the heterogeneous FPGA memories \gagan{(i.e., URAM, BRAM, and HBM)}. By using a greedy algorithm, we determine the best-suited hierarchy for our kernel. The memory controller (shown in Figure~\ref{fig:system}) handles the data placement to the appropriate memory type \gagan{based on programmer's directives}.

\gagan{On the FPGA, following the initial buffering (\circled{2}), the \gagan{transferred} grid data is mapped onto the HBM memory (\circled{3}).  As \gagan{the} FPGA ha\gagan{s} limited resources we propose a 3D window-based grid transfer from the host DRAM to the FPGA, facilitating a smaller, less power-hungry deployment. The window size represents the portion of the grid a processing element ({PE} in Figure~\ref{fig:system}) would process. \gagan{Most FPGA developers manually optimize for the right window size. However, manual optimization is tedious because of the huge design space, and \juanggg{it} requires expert guidance. 
Further, selecting an inappropriate window size lead\gagan{s} to sub-optimal results. 
\juangg{Our experiments (in Section~\ref{subsection:evaluation}) show that}: (1) finding the \juangg{best} window size is critical in terms of \juangg{the} area vs. performance trade-off, and (2) the \juangg{best window} size depends on the datatype precision. Hence, instead of pruning the design space manually, we formulate 
\juanggg{the search for} the \juangg{best} window size as a multi-objective auto-tuning problem taking into account the datatype precision. We make use of OpenTuner~\cite{opentuner}, 
\juanggg{which} uses machine-learning techniques to guide the design-space~exploration.}}


Our design consists of multiple PEs (shown in Figure~\ref{fig:system}) that exploit data-level parallelism in COSMO \gagannn{weather prediction} kernels. \gagan{A dedicated HBM memory port is assigned to a specific PE}; therefore, we enable as many HBM ports as the number of PEs. This allows us to use the high \gagan{HBM} bandwidth \gagannn{effectively} because each PE fetches from an independent port. 
In our design, we use a switch, which provides the capability to 
\juanggg{bypass} the HBM, \juangg{when the} grid size \juangg{is small}, and map the data directly onto the FPGA's URAM \gagan{and BRAM}. The HBM port provides 256-bit data, \gagannn{which} is half the size of \juangg{the} CAPI2 \gagannn{bitwidth} (512-bit). Therefore, to match \gagannn{the} CAPI2 bandwidth, we introduce a stream converter logic (\circled{4}) that converts \juangg{a} 256-bit HBM stream to \juangg{a} 512-bit stream (CAPI compatible) or vice versa.
From HBM, \gagan{a PE} reads a single stream of data that consists of all the fields\footnote{Fields represent atmospheric components like wind, pressure, velocity, etc. that are required for weather calculation.} \juangg{that} are needed for a specific COSMO kernel computation. \gagan{The PEs} use a fields stream splitter logic (\circled{5}) that splits a single HBM stream to multiple streams (512-bit each), one for each field.
  
\gagann{To optimize a PE\gagan{,} we apply various optimization strategies. First, we exploit the inherent parallelism in \gagann{a given} algorithm through hardware pipelining. \gagann{Second}, we partition on-chip memory to avoid the stalling of our pipelined design, since the on-chip BRAM/URAM has only two read/write ports. \gagann{Third}, all the tasks execute in a dataflow manner that enables task-level parallelism. \texttt{vadvc} is more \gagan{ computationally complex} than \texttt{hdiff} because it involves forward and backward sweeps \gagan{with dependencies in \juangg{the} z-dimension}. While \texttt{hdiff} performs \gagan{only} \juangg{Laplacian} and flux calculations with dependencies in \juangg{the} x- and y-dimensions. 
Therefore, we demonstrate our design flow by means of \gagannn{the} \texttt{vadvc} kernel (Figure~\ref{fig:complete_flow}). 
Note \juangg{that} we show \juangg{only} a single port-based PE operation.  \juanggg{However}, for multiple PEs, we enable multiple HBM ports.} 

We make use of memory reshaping techniques to  configure our memory space with multiple parallel BRAMs or URAMs~\cite{dioFPT}. \gagann{We form an intermediate memory hierarchy by decomposing (or slicing) 3D window data into a 2D grid. } This allows us to bridge the latency gap between the HBM memory and our accelerator. Moreover, it allows us to exploit the available FPGA resources efficiently. Unlike traditionally-fixed CPU memory hierarchies, which perform poorly with irregular access patterns and suffer from cache pollution effects, application-specific memory hierarchies \juangg{are} shown to improve \juangg{energy and latency} by tailoring the cache levels and cache sizes to \juangg{an} application's memory access patterns~\cite{jenga}.
  
The main computation pipeline (\circled{7}) \gagann{consists of a forward and a backward sweep logic}. The forward sweep \gagannn{results} are stored in \juangg{an} intermediate buffer to allow for backward sweep calculation. \juangg{Upon} completion of \juangg{the} backward sweep, results are \gagan{placed in} an output buffer \gagan{that is followed by a degridding logic (\circled{6}). The degridding logic} converts the calculated results to a 512-bit wide output stream (\circled{8}). As there is only a single output stream (both in \texttt{vadvc} and \texttt{hdiff}), we do not need extra logic to merge the streams. The 512-bit \gagan{wide} stream goes through an HBM stream converter logic (\circled{4}) that converts the stream bitwidth to HBM port size (256-bit).  
  
\gagan{Figure~\ref{fig:execution} shows the execution timeline from our host system to the FPGA board for a single PE. The host offloads the processing to an FPGA and 
\juangg{transfers} the required data \juangg{via DMA (direct memory access)} over \juangg{the} CAPI2 interface. The SNAP framework allows for parallel execution of the host and our FPGA PEs while exchanging control signals over the AXI lite interface\gagan{~\cite{axilite}}. On task completion, the AFU notif\gagannn{ies} the host system via \gagannn{the} AXI lite interface and 
\juangg{transfers} back the results \juangg{via DMA}. }

\subsection{\namePaper~Application Framework}
\begin{figure}[h]
  \centering
  \includegraphics[width=0.9\linewidth,trim={0cm 0cm 0.1cm 0cm},clip]{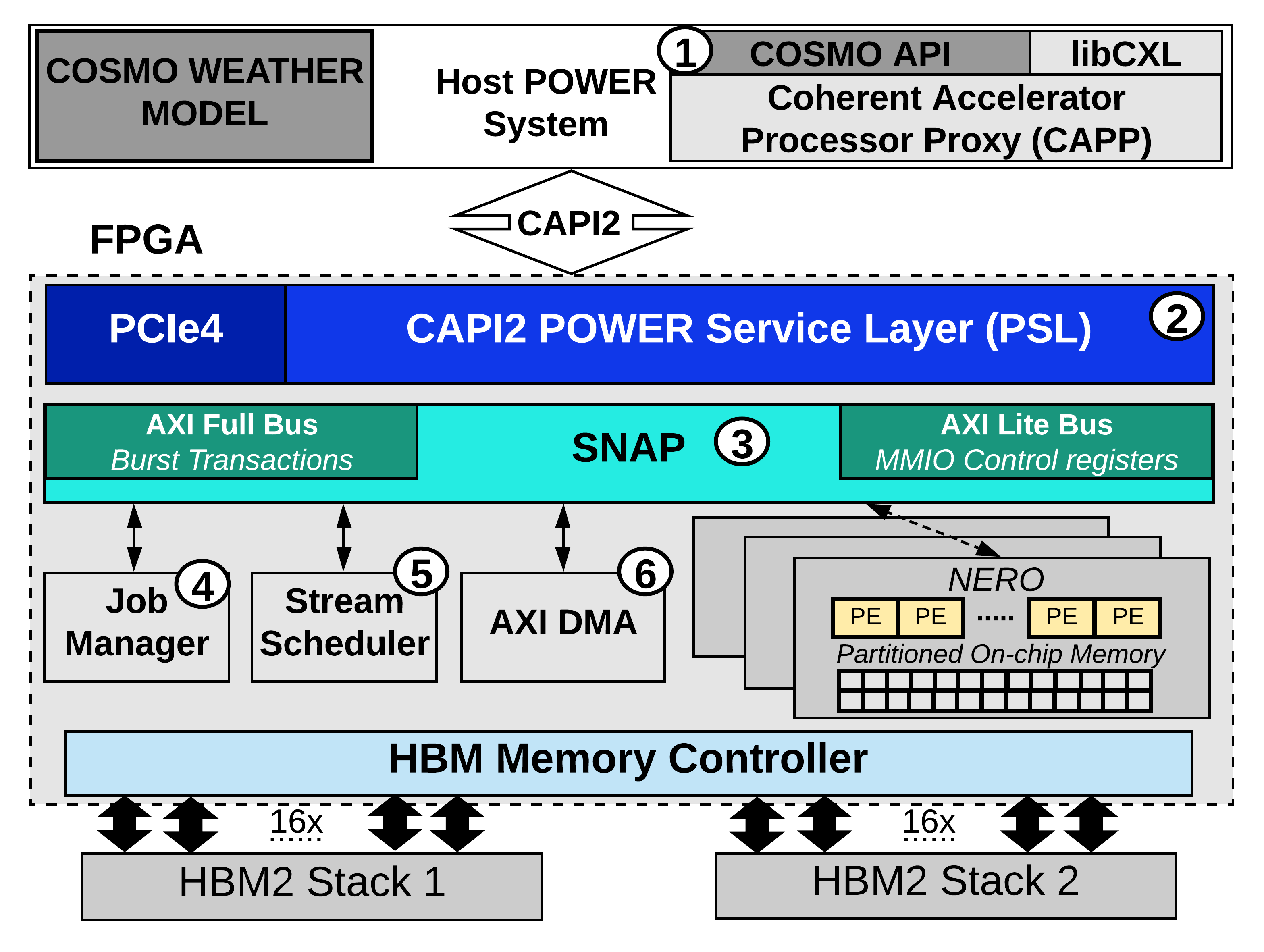} 
  \caption{
  \juangg{\namePaper~application framework.} 
  We co-design our software and hardware using \gagannn{the} SNAP framework. COSMO API allows the host to offload kernels to our FPGA platform.
 \label{fig:snap_api}}
 \end{figure}

Figure~\ref{fig:snap_api} shows the \namePaper~application framework \gagan{to support our architecture}.  A software-defined COSMO API (\circledWhite{1}) handles offloading jobs to \namePaper~with an interrupt-based queuing mechanism.  This allows for minimal CPU usage (and, hence, power \gagan{usage}) during FPGA \gagan{operation}. \namePaper~employs an array of processing elements to compute COSMO kernels, such as vertical advection or horizontal diffusion. Additionally, we pipeline our PEs to exploit the available spatial parallelism. 
By accessing the host memory through 
\juanggg{the CAPI2} cache-coherent link, \namePaper~acts as a peer to the CPU.  This is enabled \gagannn{through} \gagan{the} Power-Service Layer (PSL) (\circledWhite{2}). \gagan{ SNAP (\circledWhite{3}) allows for seamless \juangg{integration} of the COSMO API with our CAPI-based accelerator.} The job manager (\circledWhite{4}) dispatches jobs to streams, which are managed in the stream scheduler (\circledWhite{5}). The execution of a job is done by streams that determine which data is to be read from the host memory and sent to the PE array through DMA transfers (\circledWhite{6}). The pool of heterogeneous on-chip memory is used to store the input data from the main memory and the intermediate data generated by \juangg{each} PE.

%% file: results.tex
\section{Results}
\label{sec:results}

\subsection{System Integration}
We implemented our design on an Alpha-Data ADM-PCIE-9H7 card~\cite{ad9h7} featuring the Xilinx Virtex Ultrascale+ XCVU37P-FSVH2892-2-e~\cite{vu37p} and 8GiB HBM2 
\juang{(i.e., two stacks of 4GiB each)}~\cite{hbm} with an IBM POWER9 as the host system. The POWER9 socket has 16 cores, each of which supports four-thread simultaneous multi-threading. We compare our HBM-based design to a conventional DDR4 DRAM~\cite{ad9v3} based design. 
\juanggg{We perform the experiments for the DDR4-based design on} 
an Alpha-Data ADM-PCIE-9V3 card featuring the Xilinx Virtex Ultrascale+ XCVU3P-FFVC1517-2-i~\cite{vu37p}.

Table~\ref{tab:systemparameters}
provides our system parameters.
We have co-designed our hardware and software interface around the SNAP framework while using the HLS design flow.

\input{tables/system.tex}

\subsection{Evaluation}
\label{subsection:evaluation}
\begin{figure}[h]
  \centering
  \includegraphics[width=0.98\linewidth,trim={0.4cm 0.3cm 0.35cm 0.2cm},clip]{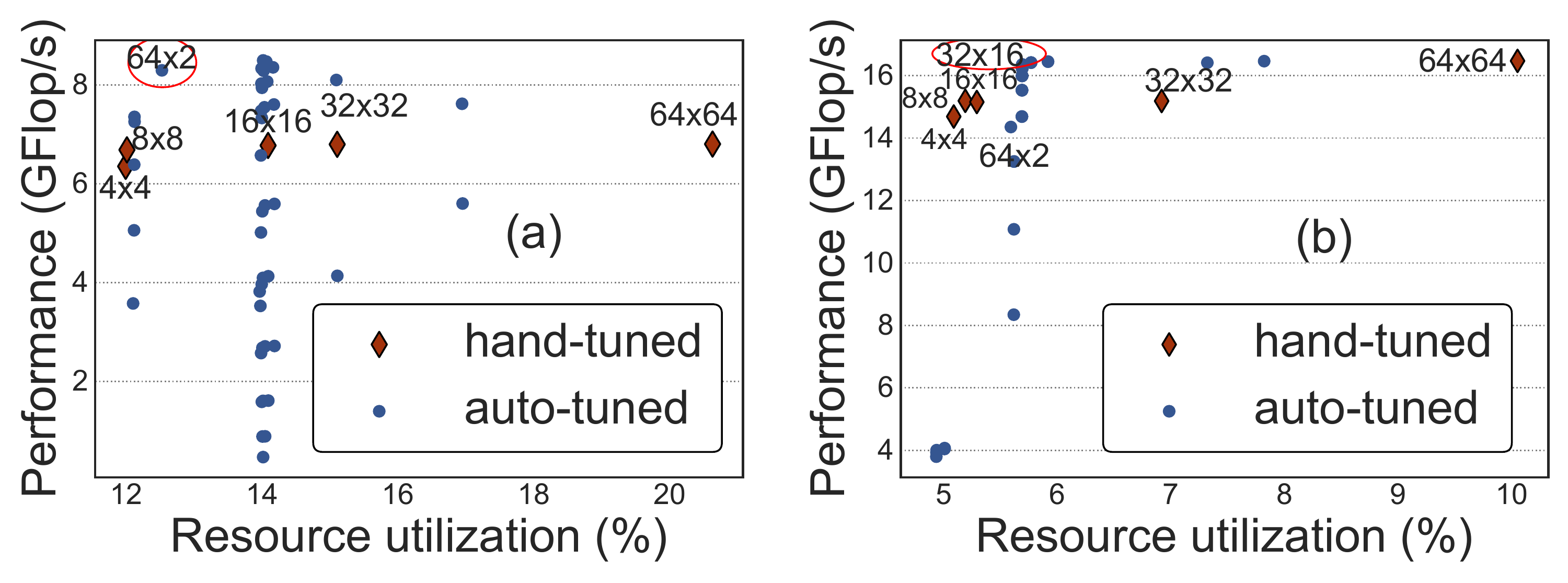}
      \caption{Performance \gagann{and FPGA resource utilization } \gagannn{of} single \texttt{vadvc} PE\gagann{, as a function of tile-size,} using hand-tuning and auto-tuning for (a) single-precision (32-bit) and (b) half-precision (16-bit). We highlight the Pareto-optimal solution that we use for our \texttt{vadvc} accelerator \gagan{(with a red circle). Note that }the Pareto-optimal solution changes with precision.
  \label{fig:single_afu}}
 \end{figure}

We 
\juanggg{run} our experiments using a $256\times256\times64$-point domain similar to the grid domain used by the COSMO \gagannn{weather prediction} model. We employ an auto-tuning technique to determine a Pareto-optimal solution (in terms of performance and resource utilization) for our 3D window dimensions.
{The auto-tuning \juanggg{with OpenTuner} exhaustively searches for every tile size in the x- and y-dimensions for \texttt{vadvc}.\footnote{\texttt{vadvc} has dependencies in \gagann{the} z-dimension; therefore, it cannot be parallelized in the z-dimension.} For \texttt{hdiff}, we consider sizes in all three dimensions. We define our auto-tuning as a multi-objective optimization with the goal \gagannn{of maximizing} performance with  minimal resource utilization.} Section~\ref{sec:design} provides further details on our design.
Figure~\ref{fig:single_afu} shows hand-tuned and auto-tuned \gagann{performance and FPGA resource utilization} results for {\texttt{vadvc}}\gagann{, as a function of the chosen tile size. From the figure, we draw two observations.} 

\gagann{First, by using the auto-tuning approach and our careful FPGA microarchitecture design, we can get \gagan{P}areto-optimal results with a tile size of $64\times2\times64$ for single-precision {\texttt{vadvc}}\gagan{, which} gives us a peak performance of 8.49 GFLOP/s. For half-precision, we use a tile size of $32\times16\times64$ to achieve a peak performance of 16.5 GFLOP/s. We employ a similar strategy for {\texttt{hdiff}} to attain a single-precision performance of 30.3 GFLOP/s with a tile size of $16\times64\times8$ and a \gagan{half-precision} performance of 77.8 GFLOP/s \gagan{with} a tile size of $64\times8\times64$. 

Second, in FPGA acceleration, designers usually rely on expert judgement to find the appropriate tile-size and often adapt the design to \gagannn{use} homogeneous tile sizes. However, as shown in Figure~\ref{fig:single_afu}\gagan{,} such hand-tuned implementations lead to sub-optimal results in terms of either resource utilization~or~performance. 

\gagann{We conclude that} the \gagann{Pareto-optimal} tile size depends on the data precision used\gagan{:} a \gagan{good} tile-size for single-precision might lead to poor results when used with half-precision.

\gagan{Figure~\ref{fig:perf} shows \gagann{single-precision} performance results for the (a) vertical advection and (b) horizontal diffusion kernels. For both kernels, we implement our design 
\juanggg{on} an HBM- and \juanggg{a} DDR4-based FPGA board. To compare the performance results, we  scale the number of PEs and analyze the change in execution time. For \gagann{the} DDR4-based design, we 
\juanggg{can accommodate only} 4 PEs on \gagann{the} 9V3 board, while for the HBM-based design, we 
\juanggg{can} fit 14 {PE}s before exhausting the on-board resources. \gagann{We draw four observations from the figure.}

 \begin{figure}[h]
  \centering
  \includegraphics[width=0.98\linewidth,trim={0.4cm 0.3cm 0.35cm 0.4cm},clip]{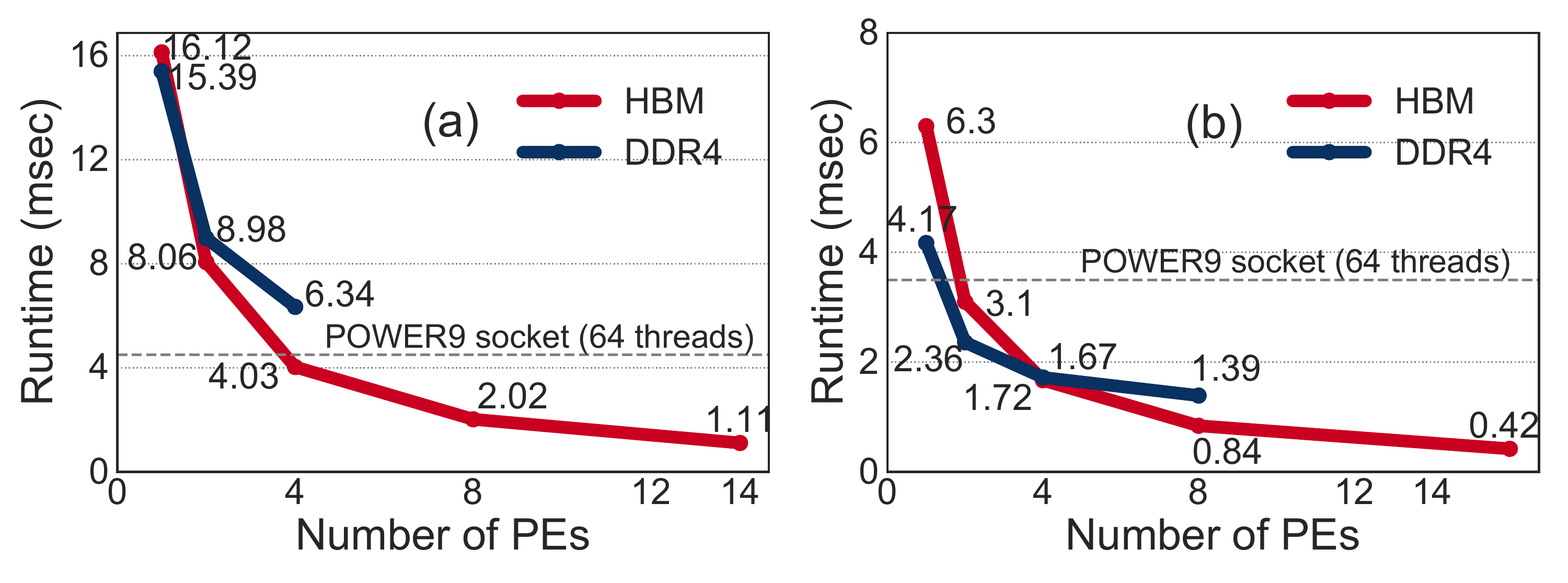}
   \caption{ \gagann{Single-precision} performance for (a) \texttt{vadvc} and (b) \texttt{hdiff}\gagann{, as a function of accelerator PE count on the HBM- and DDR4-based FPGA boards}. 
   \gagan{We also show \gagann{the}} single socket (64 threads) performance of \gagan{an} IBM POWER9 host system for both \texttt{vadvc} and \texttt{hdiff}.
  \label{fig:perf}}
 \end{figure}

First, our \gagannn{full-blown} \gagan{HBM-based} {\texttt{vadvc}} and {\texttt{hdiff}} implementations} \gagann{provide} 120.7~GFLOP/s and 485.4~GFLOP/s \gagan{performance}, which are 4.2$\times$ and 8.3$\times$ higher than the performance of a complete~POWER9~socket. 
\juanggg{For half-precision, if we use} 
the same amount of {PE}s as in single precision, \gagann{our accelerator} 
\juanggg{reaches} a performance of 247.9~GFLOP/s for \texttt{vadvc} ($2.1\times$ \gagannn{the} single-precision performance) and 1.2~TFLOP/s for \texttt{hdiff} ($2.5\times$ \gagannn{the} single-precision performance). \gagan{\gagann{Our} DDR4-based design achieves 34.1~GFLOP/s and 145.8~GFLOP/s \juanggg{for \texttt{vadvc} and {\texttt{hdiff}, respectively, which are}} 1.2$\times$ and 2.5$\times$ \juanggg{the} performance \juanggg{on the POWER9 CPU}.

Second, for a single PE, which 
\juanggg{fetches} data from a single memory channel, \gagann{the} DDR4-based design provides higher performance \juanggg{than the HBM-based design}. 
\juanggg{This is because the DDR4-based FPGA has \gagannn{a} larger bus width (512-bit) than an HBM port (256-bit). 
This \gagannn{leads to} a lower transfer rate for an HBM port (0.8-2.1 GT/s\footnote{Gigatransfers per second.}) than for a DDR4 port (2.1-4.3 GT/s). 
One way to match the DDR4 bus width would be to have a single PE fetch data from multiple HBM ports in parallel. 
However, using more ports leads to higher power consumption ($\sim$1 Watt per HBM port).}

Third, as we increase the number of PEs, we observe a linear reduction in the execution time of \gagann{the} HBM-based design. 
This is because 
we can \juanggg{evenly} divide the computation between multiple PEs\gagannn{,} each \gagan{of which} 
\juanggg{fetches data} from a separate HBM port. 

Fourth, \gagann{in the DDR4-based design, the use of only a single channel to feed multiple {PE}s leads to a congestion issue that causes a non-linear run-time reduction.} \gagann{As we \gagann{increase the number of accelerator PEs}, we observe that the} {PEs}} 
\juanggg{compete for a single memory channel, which causes frequent stalls.} \gagannn{This phenomenon leads to worse performance scaling characteristics for the DDR4-based design as compared to the HBM-based design.}


\subsection{\gagann{Energy} Analysis}
We compare the energy consumption of our accelerator to \gagan{a} \juanggg{16-core} POWER9 host system. For the POWER9 system, we use the AMESTER\footnote{https://github.com/open-power/amester} tool to measure the active power\footnote{Active power denotes the difference between the total power of a complete socket (including CPU, memory, fans, I/O, etc.) when an application is running compared to when it is idle.} consumption. 
\juanggg{We measure}~99.2~Watt\gagan{s} for \texttt{vadvc}, and ~97.9~Watt\gagan{s} for \texttt{hdiff} by monitoring \gagannn{built} power sensors in 
\juanggg{the POWER9}~system. 

{By executing these kernels on an HBM-based board, we \gagann{reduce the energy consumption by} $22\times$ for {\texttt{vadvc}} and $29\times$ for {\texttt{hdiff}} compared to 
\juanggg{the} 16-core POWER9 system.} 
\gagan{Figure~\ref{fig:energy_eff} shows the energy efficiency (GFLOPS per Watt) for {\texttt{vadvc}} and {\texttt{hdiff}} 
\juanggg{on the HBM-} and DDR4-based designs.} 
We make three \gagan{major} observations from the figure. 

 \begin{figure}[h]
  \centering
  \includegraphics[width=0.98\linewidth,trim={0.4cm 0.3cm 0.35cm 0.4cm},clip]{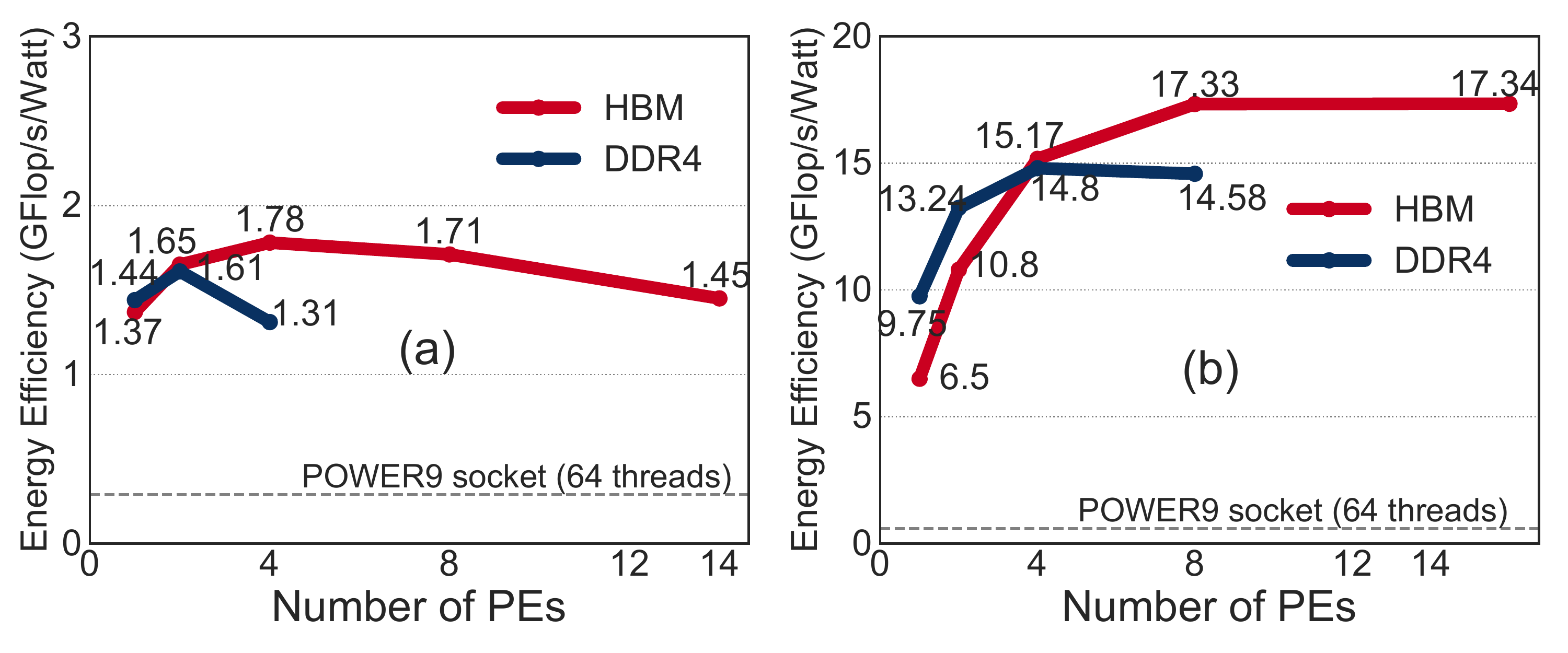}
     \caption{Energy efficiency for (a) \texttt{vadvc} and (b) \texttt{hdiff} on HBM- and DDR4-based FPGA boards. \gagann{We also show \gagann{the}} single socket (64 threads) energy efficiency of \gagan{an} IBM POWER9 host system for both \texttt{vadvc} and \texttt{hdiff}.
  \label{fig:energy_eff}}
 \end{figure}

First, with our 
\juanggg{full-blown} \gagan{HBM-based} designs \gagan{(i.e., 14 PEs for \texttt{vadvc} and 16 PEs for \texttt{hdiff})}, we achieve 
energy efficiency \juanggg{values} of 1.5~GFLOPS/Watt and 17.3~GFLOPS/Watt for {\texttt{vadvc}} and {\texttt{hdiff}}, respectively. 

Second, 
the DDR4-based \gagan{design} is more energy efficient \juanggg{than the HBM-based design} \gagannn{when the number of PEs is small}. 
\juanggg{This observation is inline with our discussion about performance with small PE counts in Section~\ref{subsection:evaluation}.} 
However, as we increase the number of {PE}s, the HBM\gagan{-based design provides better energy efficiency} for memory-bound kernels. \gagan{This is because} more data can be fetched \gagannn{and processed} in \gagan{parallel} \gagannn{via} multiple ports. 

Third, kernels like {\texttt{vadvc}}, with intricate memory access patterns, \gagan{are not able to} reach the peak \gagannn{computational power} of FPGAs. \gagann{The large amount of control flow in {\texttt{vadvc}} leads to large resource consumption. Therefore, \gagannn{when} increasing the PE count, we observe a high increase in power consumption with low energy efficiency. } 

\gagann{We conclude that} enabling many HBM ports might not always \gagannn{be} beneficial in terms of energy consumption because each HBM port \gagannn{consumes} $\sim$1 Watt of power consumption. However, \gagan{data-parallel} kernels like {\texttt{hdiff}} can achieve much higher performance in an energy efficient manner} \gagannn{with more PEs and HBM~ports.} 

\subsection{\gagann{FPGA Resource Utilization}}
Table~\ref{tab:utilization} shows the \gagan{resource} utilization \gagan{of \texttt{vadvc}} and \texttt{hdiff} on the AD9H7 board. \gagann{We draw two observations. First, there is a high BRAM consumption compared to other FPGA resources. This is because we implement input, field, and output signals as \texttt{hls::streams}. In \gagannn{high-level synthesis}, by default, streams are implemented as FIFOs that make use of BRAM. Second,  \texttt{vadvc} has a much larger resource consumption  than \texttt{hdiff} because \texttt{vadvc} has higher computational complexity and requires \gagannn{a larger} number of fields to \gagan{perform} the \gagan{compound} stencil calculation. \gagan{Note that for} \texttt{hdiff}, we can \gagan{accommodate} more {PE}s, but in this work, we  make use of only \gagan{a} single HBM stack. Therefore\gagan{,} we \gagan{use} 16 {PE}s \gagan{because} a single \gagan{HBM} stack offers 16 memory ports.}

\input{tables/utilization}

%% file: tables/system.tex
  \vspace{0.2cm}
\begin{table}[h]
  \caption{\gagan{S}ystem parameters and hardware configuration for the CPU and the FPGA board.}
  \vspace{-0.4cm}
    \label{tab:systemparameters}
      \begin{center}
      \small
\resizebox{\linewidth}{!}{%
\begin{tabular}{|l@{\hspace{0.1\tabcolsep}}|p{5cm}|}

\hline
\textbf{Host CPU} 

 & 16-core IBM POWER9 AC922 \\&@3.2 GHz, 4\gagann{-way} SMT\\
 \hline
 \textbf{Cache-Hierarchy}&32 KiB L1-I/D, 256 KiB L2, 10 MiB L3 \\
 \hline
 \textbf{System Memory}&32GiB RDIMM DDR4 2666 MHz \\
\hline

 \begin{tabular}[c]{@{}l@{}}
 \textbf{HBM-based} \\\textbf{FPGA Board} \end{tabular}  &
  \begin{tabular}[c]{@{}l@{}}
 Alpha Data
ADM-PCIE-9H7\\
 Xilinx Virtex Ultrascale+ XCVU37P-2\\
  8GiB (HBM2) with PCIe Gen4 x8\\
 \end{tabular}\\
\hline
 \begin{tabular}[c]{@{}l@{}}
 \textbf{DDR4-based} \\\textbf{FPGA Board} \end{tabular}  & \begin{tabular}[c]{@{}l@{}}
 Alpha Data
ADM-PCIE-9V3\\
 Xilinx Virtex Ultrascale+ XCVU3P-2\\
 8GiB (DDR4) with PCIe Gen4 x8
  \end{tabular}\\



\hline

\end{tabular}
}
  \end{center}
 \end{table}

%% file: tables/utilization.tex
  \vspace{0.2cm}
\begin{table}[h]
  \caption{FPGA resource utilization \gagannn{in} \gagan{our highest-performing HBM-based design\gagannn{s for}} \texttt{vadvc} and \texttt{hdiff}. }
  \vspace{-0.4cm}
    \label{tab:utilization}
          \begin{center}
\resizebox{\linewidth}{!}{%
\begin{tabular}{llllllc}
\hline
\textbf{Algorithm} & \textbf{BRAM} & \textbf{DSP} & \textbf{FF} & \textbf{LUT} & \textbf{URAM} \\ \hline
\texttt{vadvc}              & 81\%            & 39\%           & 37\%          & 55\%           & 53\%                         \\ 
\texttt{hdiff}              & 58\%            & 4\%            & 6\%           & 11\%           & 8\%                           \\ \hline
\end{tabular}
}
  \end{center}
\vspace{-0.2cm}
\end{table}

%% file: relatedWork.tex
\section{Related Work}
\label{sec:relatedWork}
\juang{To our knowledge, this is the first work to evaluate the benefits of using FPGAs equipped with \gagann{high-bandwidth memory (HBM) to accelerate} stencil computation.} \gagannn{We exploit near-memory capabilities of such FPGAs to accelerate important weather prediction kernels.}

\gagannn{Modern workloads exhibit limited locality and operate on large amounts of data, which causes frequent data movement between the memory subsystem and the processing units~\cite{ghose2019processing,mutlu2019,googleWorkloads,mutlu2019enabling}. This frequent data movement has a severe impact on overall system performance and energy efficiency. 
A way to alleviate this \emph{data movement bottleneck}~\cite{singh2019near, mutlu2019, ghose2019processing,mutlu2019enabling,googleWorkloads} is \emph{near-memory computing} (NMC), which consists of placing processing units closer to memory. 
NMC is enabled by new memory technologies, such as 3D-stacked memories~\cite{7477494,6757501,6025219,hbm,lee2016smla}, and also by cache-coherent interconnects~\cite{openCAPI, benton2017ccix, sharma2019compute},  which allow close integration of processing units and memory units. 
Depending on the applications \gagannn{and systems} of interest (e.g.,~\cite{nai2017graphpim,7056040,lee2018application,kang2013enabling,hashemi2016continuous,akin2015data,babarinsa2015jafar,lee2015bssync,chi2016prime,kim2016neurocube,asghari2016chameleon,boroumand2016lazypim,seshadri2015gather,liu2017concurrent,gao2015practical,morad2015gp,googleWorkloads,teserract,ahn2015pim,hsieh2016accelerating,hashemi2016accelerating}), prior works propose different types of near-memory processing units, such as general-purpose CPU cores~\cite{lee2018application,alian2018application,de2017mondrian,koo2017summarizer,7927081,teserract,nair2015active,6844483,googleWorkloads,boroumand2016lazypim,boroumand2019conda}, GPU cores~\cite{zhang2014top,7756764,7551394,ghose2019demystifying}, reconfigurable units~\cite{7446059, jun2015bluedbm,istvan2017caribou,narmada}, or fixed-function units~\cite{ahn2015pim,hsieh2016accelerating,gu2016biscuit,nai2017graphpim,liu2018processing,kim2018grim,hashemi2016accelerating,hashemi2016continuous}}.

FPGA accelerators are promising to enhance overall system performance \gagann{with low power consumption.}
Past works~\cite{jun2015bluedbm, kara2017fpga, alser2019shouji, giefers2015accelerating,8373077,Lee:2017:EBG:3137765.3137776,alser2017,chai_icpe19,chang2017collaborative,jiang2020,alser2019sneakysnake} show that FPGAs can be employed {effectively} for a wide range of applications. The recent addition of HBM to \juang{FPGAs} 
\juang{presents an opportunity to exploit} 
high \juang{memory} bandwidth with \gagannn{the} low-power FPGA fabric. 
The potential of high-bandwidth memory~\cite{hbm,lee2016smla} has been explored in many-core processors~\cite{hbm_joins,ghose2019demystifying} and GPUs~\cite{hbm_gpu_data_intensive,ghose2019demystifying}. 
\juang{A recent work~\cite{wang2020} shows the potential of HBM for FPGAs with a memory benchmarking tool.}
\juang{\namePaper} is the first work to accelerate a real-world HPC weather prediction application using \juang{the} FPGA+HBM fabric. {Compared to a previous work~\cite{narmada} that \juang{optimizes only the} horizontal diffusion kernel \juang{on an FPGA with DDR4 memory}, our analysis reveals \juang{that the} vertical advection kernel has a much lower compute intensity with little to no regularity.
Therefore, our work \juang{accelerates} both kernels that together represent the algorithmic diversity of the entire COSMO \gagannn{weather prediction} model. Moreover, compared to 
\cite{narmada}, \juang{\namePaper~\gagannn{improves performance by}} $1.2\times$ on a DDR4\gagannn{-}based \gagann{board} and $37\times$ on an HBM-based board for horizontal diffusion by using a dataflow implementation with auto-tuning.}

\gagan{Enabling higher performance for stencil computations has been a subject of optimizations 
across the \juang{whole} computing stack~\cite{sano2014multi,7582502,chi2018soda,de2018designing,christen2011patus,  datta2009optimization, meng2011performance, henretty2011data, strzodka2010cache, tang2011pochoir,gonzalez1997speculative,armejach2018stencil,gysi2015modesto}}. \juang{Szustak~\etal} accelerate the MPDATA advection scheme on multi-core CPU~\cite{szustak2013using} and computational fluid dynamics kernels on FPGA~\cite{mpdata}. Bianco~\etal~\cite{bianco2013gpu} \juan{optimize the COSMO \gagannn{weather prediction} model} for GPUs while Thaler~\etal~\cite{cosmo_knl} \juan{port} COSMO to a many-core system.  Wahib~\etal~\cite{wahib2014scalable} \juang{develop} an analytical performance model for choosing an optimal GPU-based execution strategy for various scientific applications, including COSMO. 
Gysi~\etal~\cite{gysi2015modesto} \juang{provide} guidelines for optimizing stencil kernels for CPU--GPU systems.


%% file: conclusion.tex
\section{Conclusion}
\label{sec:conclusion}
\juang{We introduce} \namePaper, the first design and implementation on a reconfigurable fabric \juang{with \gagann{high-bandwidth memory} (HBM)} to accelerate representative weather prediction kernels, i.e., vertical advection (\texttt{vadvc}) and horizontal diffusion (\texttt{hdiff}), from a real-world weather prediction application. 
These kernels are compound stencils that are found in various weather prediction applications, including the COSMO  model. \gagannn{We show that c}ompound kernels do not perform well on conventional architectures due to their complex data access patterns and low data reusability, which \gagann{make} them memory-bounded. 
\juang{Therefore, they greatly benefit from our near-memory computing solution \gagannn{that} takes advantage of the high \gagann{data transfer} bandwidth of HBM.}


\juang{\namePaper's implementations of \texttt{vadvc} and \texttt{hdiff} outperform the optimized software implementations on a 16-core POWER9 with \gagann{4-way multithreading} by $4.2\times$ and $8.3\times$, with $22\times$ and $29\times$ less energy consumption, respectively. 
We conclude that hardware acceleration on \gagann{an} FPGA+HBM fabric is a promising solution for compound stencil\gagann{s} present in weather prediction applications. \gagann{We hope that o}ur \gagann{reconfigurable near\gagannn{-}memory} accelerator inspire\gagann{s} developers of different high-performance computing applications that suffer from \gagann{the} memory bottleneck.
}